\documentclass[a4paper]{PoS}
\pdfoutput=1
\usepackage{amsmath,bbm}

\let\OLDthebibliography\thebibliography
\renewcommand\thebibliography[1]{
  \OLDthebibliography{#1}
  \setlength{\parskip}{0pt}
  \setlength{\itemsep}{1.2mm}
}

\newcommand{\bphi}{\bar{\phi}}

\newcommand{\BZ}{_{\rm BZ}}

\newcommand{\Tr}{{\rm Tr}}
\newcommand{\SU}{{\rm SU}}
\newcommand{\U}{{\rm U}}

\newcommand{\I}{\mathcal{I}}

\newcommand\Nb{N_b}
\newcommand\Nc{N_c}

\newcommand\mb{m}

\newcommand\1{\mathbbm{1}}
\newcommand\C{\mathbbm{C}}
\newcommand\vp{\mathbf{p}}

\renewcommand\phi\varphi

\title{Induced QCD with two auxiliary bosonic fields\thanks{Work
    supported in part by DFG in the framework of SFB/TRR-55.}}

\ShortTitle{Induced QCD with two auxiliary bosonic fields}

\author{\speaker{Bastian B. Brandt} and Tilo Wettig\\
  Institute for Theoretical Physics\\
  University of Regensburg \\ D-93040 Regensburg \\
  E-mail: \email{bastian.brandt@ur.de}, \email{tilo.wettig@ur.de}}

\abstract{Following a proposal of Budczies and Zirnbauer, we
  investigate an alternative lattice discretization of continuum
  $\SU(\Nc)$ Yang-Mills theory in which the self-interactions of the
  gauge field are induced by a path integral over $\Nb\ge\Nc-1$
  auxiliary bosonic fields which are coupled linearly to the gauge
  field.  In two dimensions there exists an analytic proof that the
  new discretization reproduces Yang-Mills theory in its
  non-perturbative continuum limit.  We provide numerical evidence
  that this is also the case in three and four dimensions and that,
  after a suitable matching of the free parameters, the results of the
  induced theory agree with results from the ordinary plaquette action
  up to lattice artifacts.  The new discretization is ideally suited
  to change the order of integration in the QCD path integral to
  arrive at formulations in which the gauge fields have been
  integrated out.  The resulting theories might be amenable to methods
  previously used in the infinite-coupling limit, and we briefly
  discuss possibilities to arrive at dual representations of lattice
  QCD.}

\FullConference{The 32nd International Symposium on Lattice Field Theory\\
         23-28 June, 2014\\
         Columbia University New York, NY}

\begin{document}

\section{Introduction}

Strong-coupling approaches to lattice QCD allow for analytical studies
and the construction of new simulation algorithms. Typically, these
methods are applicable only if the self-interactions of the gluons are
neglected, i.e., in the limit of infinite coupling. There have been
several attempts to overcome this problem by ``inducing'' the gluon
dynamics with an action formulated in terms of auxiliary
fields~\cite{Bander:1983mg,Hamber:1983nm,Kazakov:1992ym,Hasenfratz:1992jv}. One
of the typical problems for these theories of an ``induced'' version
of Yang-Mills theory is that Yang-Mills theory is usually recovered in
the limit of an infinite number of auxiliary fields,\footnote{An
  exception is the Kazakov-Migdal model, which, however, does not
  reproduce Yang-Mills theory.} rendering the application of the
methods impractical.

In~\cite{Budczies:2003za} Budczies and Zirnbauer (BZ) proposed a
``designer action'' (in their paper for gauge group $G=\U(\Nc)$)
conjectured to reproduce continuum Yang-Mills theory when the mass of
the auxiliary boson fields approaches a critical value. The major
novelty is that this works already for a fixed number of auxiliary
fields $\Nb\ge\Nc$ for $G=\U(\Nc)$ (or $\Nb\ge \Nc-1$ for
$G=\SU(\Nc)$, see below). This was shown analytically for $G=\U(\Nc)$
in $d=2$ dimensions by matching to an earlier result of
Witten~\cite{Witten:1991we}. For $d>2$ there is no analytical proof
but a plausible universality argument.

In this proceedings article we investigate the properties of a
modified version of the BZ theory for $\SU(\Nc)$ gauge theory and
provide numerical evidence that the continuum limit of the theory
equals continuum Yang-Mills theory.  The modified version solves a
sign problem of the bosonized version of the theory. Here we restrict
ourselves to showing only a few numerical and analytical results. More
details will be given in a future publication~\cite{future-paper}. In
section~\ref{sec:dual-rep} we briefly discuss how this theory can be
used to arrive at dual representations of full QCD.

\section{Induced Yang-Mills theory}
\label{sec:bz-theory}

The action introduced by Budzcies and Zirnbauer
in~\cite{Budczies:2003za} is given for a hypercubic lattice by
\begin{equation}
  \label{eq:BZ-action}
  S\BZ [\phi,\bphi,U] = \sum_{b=1}^{\Nb} \sum_{\pm \vp}
  \sum_{j=1}^{4} \big[ m\BZ \bphi_{b,\vp}(x_j^\vp)\phi_{b,\vp}(x_j^\vp) -
  \bphi_{b,\vp}(x_{j+1}^\vp) U(x_{j+1}^\vp,x_j^\vp) \phi_{b,\vp}(x_j^\vp)
  \big] \,,
\end{equation}
where $U\in G$ are the link variables, $\phi$ are auxiliary bosonic
fields, $b$ labels the number of boson flavors, and $m\BZ$ is a mass
parameter (which we take to be real) satisfying $m\BZ>1$. The index
$\vp$ labels oriented plaquettes, and $j$ labels the points of the
plaquette in order of the orientation. Integration over the auxiliary
fields in the path integral yields the weight factor
\begin{equation}
 \label{eq:BZ-det-action}
  \omega\BZ[U] = \int [d\phi] \exp\big\{-S\BZ [\phi,\bphi,U]\big\} = \prod_{p}
\big| \det\big(m\BZ^4-U_p\big) \big|^{-2\Nb} \,,
\end{equation}
where the product is over all unoriented plaquettes $p$, i.e., every
plaquette is counted only once, and $U_p$ is the usual product of
$U$-fields around the plaquette. We rewrite this factor in terms of the
coupling $\alpha\BZ=m\BZ^{-4}$ as
\begin{equation}
 \label{eq:BZ-det-action2}
  \omega\BZ[U] \sim \prod_{p} \big| \det\big(1-\alpha\BZ\,U_p\big) \big|^{-2\Nb}
\,.
\end{equation}

As shown in~\cite{Budczies:2003za} for $G=\U(\Nc)$, this theory has a
continuum limit for a fixed number of bosonic fields $\Nb$ when
$\alpha\BZ\to1$ as long as $\Nb\geq\Nc$. This can be shown by
investigating the behavior of the weight factor for a particular
plaquette $U_p$ in this limit.  The continuum limit is obtained when
the weight factor approaches the $\delta$-function, i.e., for any
analytic function $f:G \to \C$
\begin{equation}
 \label{eq:dfunc-lim}
  \lim_{\alpha\BZ\to1} \frac{\langle f\rangle}{\langle1\rangle} =
f(\1) \,, \quad \langle f\rangle = \int_{G} dU_p \,
f(U_p) \big| \det\big(1-\alpha\BZ U_p\big) \big|^{-2\Nb} \,.
\end{equation}
Furthermore, for the special case of $d=2$ the continuum limit
reproduces the boundary partition function of Wittens combinatorial
treatment of Yang-Mills theory~\cite{Witten:1991we}. For $d>2$ there
is no strict proof, but it was argued in~\cite{Budczies:2003za} that
in higher dimensions the collectivity of the gauge fields is
increased, which works in favor of ``universality''.  Thus, if
Yang-Mills theory is induced in $d=2$, it should also be induced in
$d>2$.

\begin{figure}[t]
\vspace*{-3mm}
\begin{minipage}[c]{.48\textwidth}
\centering
\includegraphics[width=.95\textwidth]{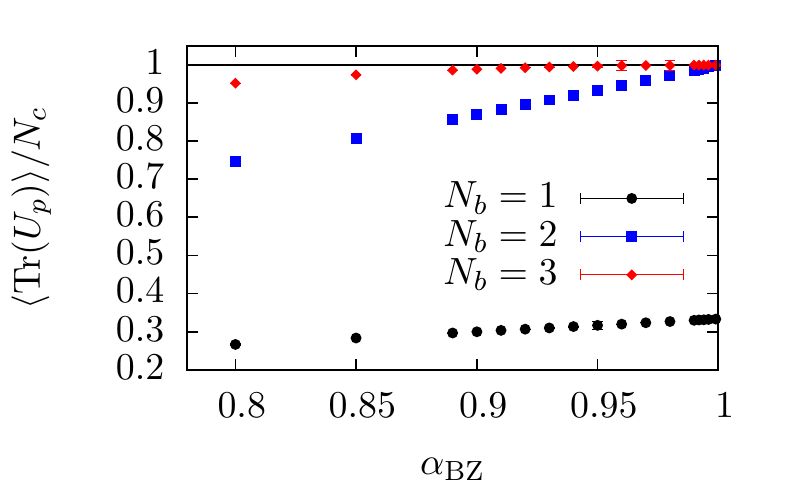}
\end{minipage}
\begin{minipage}[c]{.48\textwidth}
\centering
\includegraphics[width=.95\textwidth]{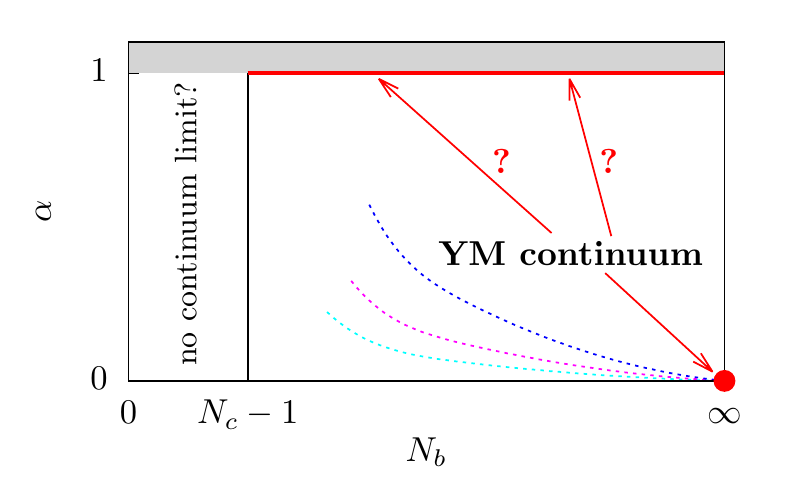}
\end{minipage}
\vspace*{-2mm}
\caption{{\bf Left:} Expectation value \protect\eqref{eq:dfunc-lim} of
  the test function $f(U_p)=\Tr(U_p)/\Nc$ for $G=\SU(3)$ in the
  approach to the limit $\alpha\BZ=1$.  {\bf Right:} Conjectured phase
  diagram in the $(\Nb,\alpha)$-plane for $G=\SU(\Nc)$. The dashed
  lines indicate possible lines of constant physics.}
\label{fig:phdiag}
\end{figure}

The extension of the existence proof of a continuum limit from
$G=\U(\Nc)$ to $G=\SU(\Nc)$ is not completely straightforward.  The
exception is the $\SU(2)$ case for which a proof with $\Nb=1$ was
already given in~\cite{Budczies:2003za}. We will discuss the details
of the proof for $G=\SU(\Nc)$ in our future
publication~\cite{future-paper}. Instead of considering the integral
analytically one can also investigate the behavior of the one-link
integral \eqref{eq:dfunc-lim} for certain test functions numerically
in the limit $\alpha\BZ\to1$. The results for $G=\SU(3)$ and
$f(U_p)=\Tr(U_p)/\Nc$ for different values of $\Nb$ are shown in
figure~\ref{fig:phdiag} (left). The plot indicates that the integral
indeed approaches $f(\1)=1$ once we have reached the critical number
of boson fields, which appears to be $\Nb=\Nc-1$ for $\SU(\Nc)$ in
contrast to $\Nb=\Nc$ for $\U(\Nc)$. Other test functions show the
same behavior. This implies that for $G=\SU(\Nc)$ the continuum limit
exists for $\alpha\BZ\to1$ if $\Nb\geq\Nc-1$. This is consistent with
our analytical findings.\footnote{Here we consider only integer values
  of $\Nb$. Non-integer values will be discussed
  in~\cite{future-paper}.}

\section{Sign problem and modified theory}
\label{sec:mod-theory}

While the weight factor \eqref{eq:BZ-det-action2} is positive
definite, this is not the case for the exponential of the
action~\eqref{eq:BZ-action} considered in~\cite{Budczies:2003za},
which is complex for a general configuration of auxiliary bosonic
fields. In future applications, e.g., simulations of the bosonized
version, this would be problematic. Fortunately, the sign problem can
be eliminated by a simple reformulation of the theory. The
corresponding weight factor is given by
\begin{equation}
  \label{eq:det-action}
  \omega[U] = \prod_{p}
  \Big[\det\big(1-\frac{\alpha}{2}\big(U_p+U_p^\dagger\big)\big)\Big]^{-\Nb} \,,
\end{equation}
where $\alpha$ is related to $\alpha\BZ$. The equivalence of the two
weight factors up to an unimportant constant is evident when the
absolute value in \eqref{eq:BZ-det-action2} is written out
explicitly. The corresponding action with auxiliary bosonic fields is
given by 
\begin{equation}
  \label{eq:action}
  \begin{array}{rcl}
  \displaystyle S_B [\phi,\bphi,U] &= \displaystyle \sum_{b=1}^{\Nb} \sum_{p}
\sum_{j=1}^{4}   \Big[ &\mb\bphi_{b,p}(x^p_j)\phi_{b,p}(x^p_j) -
\bphi_{b,p}(x^p_{j+1})   U(x^p_{j+1},x^p_j)\phi_{b,p}(x^p_j) \Big.
\vspace*{2mm}\\ & & \displaystyle \Big. - \bphi_{b,p}(x^p_{j})
U(x^p_j,x^p_{j+1}) \phi_{b,p}(x^p_{j+1})\Big]
  \end{array}
\end{equation}
with $2/\alpha=m^4-4m^2+2$ and $x_j^p=x_j^{+\vp}$. Note that $\alpha\BZ\to1$ is
equivalent to $\alpha\to1$. In figure~\ref{fig:phdiag} (right) we show the
conjectured phase diagram in the $(\Nb,\alpha)$-plane for $G=\SU(\Nc)$.

\section{Numerical investigation of the continuum limit}

We now investigate the properties of the limit $\alpha\to1$ in the
numerically cheap case of three-dimensional $\SU(2)$ gauge theory via
simulations with the weight factor~\eqref{eq:det-action}.\footnote{In
  a future publication we will also show results for $\SU(3)$ in four
  dimensions.} We fix $\Nb$ to 1 and 2 to see whether both cases lead to the
correct continuum limit. To make
contact with Yang-Mills theory we compare the simulation results to
results obtained with the standard Wilson plaquette action. The
simulations for the induced theory are done using local Metropolis
updates and links that evolve randomly in an $\epsilon$-ball around
the old ones. For the Wilson action we use the standard mixture of
heatbath and overrelaxation updates. In pure gauge theory the only
quantity that has to be fixed to render the simulations predictive is
the lattice spacing $a$, here determined via the Sommer scale
$r_0$~\cite{Sommer:1993ce}. If both theories approach the same
continuum theory, one expects that below some critical lattice spacing
both theories give the same results up to lattice artifacts.

We observe that for the whole range of couplings $\beta$ typically
considered in simulations of $\SU(2)$ gauge theory at $d=3$ (see,
e.g., \cite{Brandt:2009tc}) there exists a matching coupling $\alpha$
with the same value of $r_0/a$.  Using this as a matching condition,
we obtain a matching relation of the form
\begin{equation}
 \beta(\alpha)= \frac{b_{-1}}{1-\alpha} + b_0 + b_1(1-\alpha) + \ldots \,.
\end{equation}
This relation is valid for the range of lattice spacings used in our
simulations and all values of $\Nb\geq\Nc-1$ with $\Nb$-dependent
coefficients $b_n$.\footnote{We have explicitly tested this relation
  for $1\leq\Nb\leq5$ and for some non-integer values of $\Nb$.} In
particular, the expected divergence at $\alpha=1$ behaves like
$1/(1-\alpha)$ to very good accuracy.

\begin{figure}[t]
\vspace*{-3mm}
\begin{minipage}[c]{.48\textwidth}
\centering
\includegraphics[width=.95\textwidth]{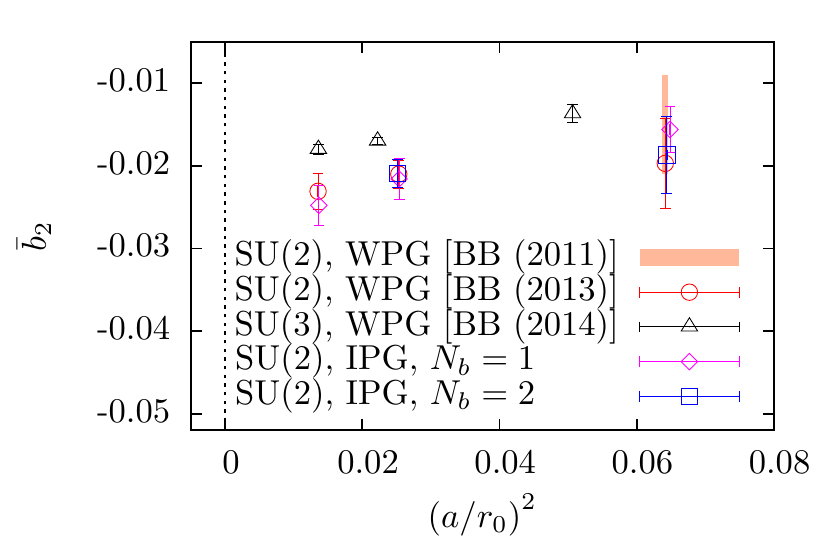}
\end{minipage}
\begin{minipage}[c]{.48\textwidth}
\centering
\includegraphics[width=.95\textwidth]{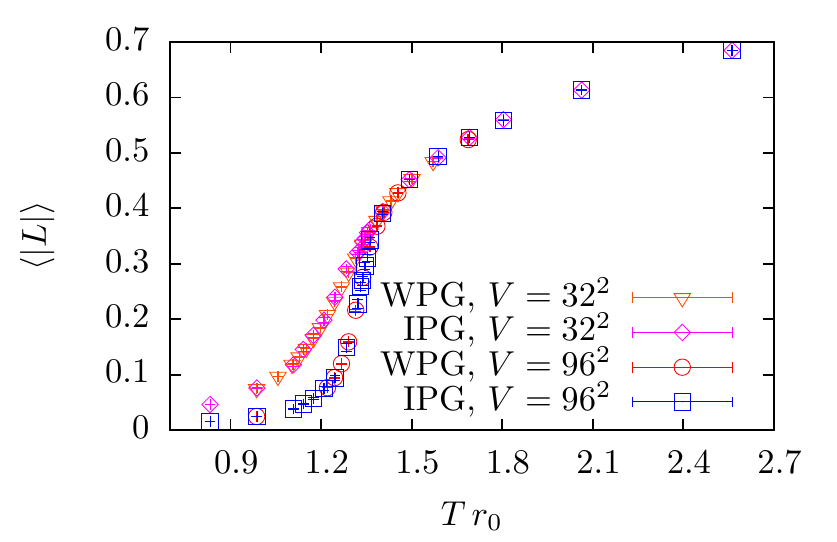}
\end{minipage}
\vspace*{-2mm}
\caption{{\bf Left:} Comparison of results from the induced (IPG) and
  the Wilson (WPG) action~\cite{Brandt:2013eua,Brandt:2010bw} for
  $\bar{b}_2$ vs.  $a^2$. {\bf Right:} Results for the absolute value
  of the Polyakov loop on the two extremal volumes at $N_t=4$.}
\label{fig:b2-extr}
\end{figure}

Using this matching and tuning the couplings to the ones
from~\cite{Brandt:2013eua} we have performed high-precision
simulations in the induced theory for the static $q\bar{q}$-potential
via Polyakov loop correlators evaluated with the multilevel
algorithm~\cite{Luscher:2001up}. We have then reproduced the analysis
of~\cite{Brandt:2013eua} and extracted the boundary coefficient
$\bar{b}_2$~\cite{Aharony:2010db} appearing in the effective string
theory for the QCD flux tube (for a review see, e.g.,
\cite{Lucini:2012gg}). Note that our focus here is on a like-by-like
comparison and not on validating the string picture. The quantity
$\bar{b}_2$ is ideal in this respect since it probes the subleading
properties of the potential and, presumably, is non-universal. The
details of the analysis, the computations, and a full list of
references will be given in~\cite{future-paper}. The results for
$\bar{b}_2$ are shown in figure~\ref{fig:b2-extr} (left) for different
values of $r_0$ in comparison to the results
of~\cite{Brandt:2013eua,Brandt:2010bw}. The plot indicates good
agreement of the results with the ones obtained with the Wilson
action. Furthermore, the results are significantly different from the
ones obtained in $\SU(3)$ and $Z_2$ gauge theory~\cite{Billo:2012da},
leading to the conclusion that the continuum potential indeed
corresponds to that of $\SU(2)$ Yang-Mills theory.

We have also investigated the finite-temperature transition for
$N_t=4$ with volumes between $V/a^2=32^2$ and $96^2$. The Polyakov
loop on the two extremal volumes with $\Nb=2$ is shown in
figure~\ref{fig:b2-extr} (right).  The plot shows very good agreement
between the two sets of points.  Performing an extraction of critical
exponents following \cite{Engels:1996dz} we obtain $T_c r_0=1.34(2)$
and $\gamma/\nu=1.69(4)$ in good agreement with the corresponding
Wilson results. To confirm this agreement we will increase the
statistics at $T_c$ and repeat the analysis for $N_t=6$.  Results for
$\Nb=1$ will be shown in \cite{future-paper}.

\section{Dual representations of QCD}
\label{sec:dual-rep}

A very interesting feature of the induced theory is that it allows us
to change the order of integration in the QCD path integral, i.e., to
integrate over the gauge fields first. The part of the
action~\eqref{eq:action} containing the link $U_\mu(x)$ (here we have
switched to the standard notation where $x$ is a particular point on
the lattice) can be written as
\begin{equation}
S_{B;x,\mu}^\text{hop}(\phi,\bphi,U)
  = - \frac{1}{2} \Tr\left[ A_\mu(x,\phi,\bphi) U 
  + A^\dagger_\mu(x,\phi,\bphi) U^\dagger \right] \,,
\end{equation}
where $A_\mu$ is a complex matrix constructed from the auxiliary
fields. Since the weight for a given link does not depend on other
link variables the path integral over the gluonic degrees of freedom
factorizes into a product over integrals of the type
\begin{equation}
 \label{eq:I-int}
 \I_\mu(x,\phi,\bphi) = \int_G dU \, e^{\frac{1}{2} \Tr[
    A_\mu(x,\phi,\bphi) U + A^\dagger_\mu(x,\phi,\bphi) U^\dagger]} \,.
\end{equation}
The exact solution of this integral is known for
$G=U(\Nc)$~\cite{Brower:1980rp} and for some $\Nc$ also for
$G=\SU(\Nc)$~\cite{Creutz:1978ub,Brower:1980tb,Eriksson:1980rq}. In
our future publication \cite{future-paper} we will derive a general
solution for $G=\SU(\Nc)$,
\begin{equation}
  \label{eq:I-solve}
  \I = 2^{\Nc(\Nc-1)/2}\biggl(\prod_{n=1}^{\Nc-1}n!\biggr)
  \sum_{\nu=0}^\infty\varepsilon_\nu\cos\big(\nu\theta\big)
  \frac{\det\big[\lambda_i^{j-1}
    I_{\nu+j-1}\big(\lambda_i\big)\big]}
  { \Delta\big(\lambda^2\big)} \,,
\end{equation}
where we have suppressed the dependence on $\mu,x,\phi$, and $\bphi$
for simplicity.  $\varepsilon_\nu$ is Neumann's factor
($\varepsilon_{\nu=0}=1$ and $\varepsilon_{\nu>0}=2$), $\exp(i\theta)$
is the complex phase of $\det(A)$, the $\lambda_i^2$ are the $\Nc$
eigenvalues of $AA^\dagger$, $\Delta$ is the Vandermonde determinant,
and $I$ denotes modified Bessel functions of the first kind.

Using this result the dual partition function of the induced theory is
given by
\begin{equation}
 \label{eq:dual-pfunc-sun}
   Z = \int [d\phi]\, \exp\Big\{-\sum_{b=1}^{\Nb} \sum_{p}
\sum_{j=1}^{4} \mb\bphi_{b,p}(x^p_j)\phi_{b,p}(x^p_j)\Big\} \prod_{x,\mu}
\I_\mu(x,\phi,\bphi) \,.
\end{equation}
Another dual partition function with a similar structure has been
obtained for the Wilson action using Hubbard-Stratonovich
transformations~\cite{Vairinhos:2014uxa}.

Full QCD also includes dynamical fermions. Those can be added to the
induced theory in the standard way using the lattice discretization of
choice. The partition function for full QCD is then
\begin{equation}
 \label{eq:QCD-pint}
 Z = \int [dU] \, [d\phi] \, [d\psi] [d\bar{\psi}]
\exp\big\{-S_F[\psi,\bar{\psi},U] -S_B [\phi,\bphi,U] \big\} \,.
\end{equation}
Using staggered-type fermions, possibly at nonzero chemical potential,
the fermion action is
\begin{equation}
 \label{eq:stagg-action}
 S_F = \sum_{x} \Big\{ \sum_\mu \left[ \bar{\psi}(x) \rho_\mu(x)
\, U_\mu(x) \psi(x+\hat{\mu}) + \bar{\psi}(x+\hat{\mu}) \tilde{\rho}_\mu(x) \,
U^\dagger_\mu(x) \psi(x) \right] + m_q \bar{\psi}(x) \psi(x) \Big\} \,,
\end{equation}
where the factors $\rho$ and $\tilde{\rho}$ include the usual
staggered phases and factors originating from the introduction of the
chemical potential.  There are several options how to proceede to
derive dual representations of QCD.  One possibility is to expand the
exponential over Grassmann fields as
\begin{equation}
 \label{eq:ferm-act-exp}
 e^{-S_F} 
= \prod_x \Big\{ \Big[ \prod_{a} \sum_{n_a(x)}
\big( M_a(x) \big)^{n_a(x)} \Big] \prod_\mu \prod_{a,b} \Big[
\sum_{p_{ab}(x,\mu)} \big( H_{ab,\mu}(x) \big)^{p_{ab}(x,\mu)}
\sum_{\tilde{p}_{ab}(x,\mu)} \big( \widetilde{H}_{ab,\mu}(x)
\big)^{\tilde{p}_{ab}(x,\mu)} \Big] \Big\} 
\end{equation}
with $\:\:M_a(x) = - m_q \; \bar{\psi}_a(x) \psi_a(x)$,
$\:\:H_{ab,\mu}(x) = - \rho_\mu(x) \, [U_\mu(x)]_{ab} \,
\bar{\psi}_a(x) \psi_b(x+\hat{\mu})\:\:$ and
$\:\:\widetilde{H}_{ab,\mu}(x) = \\ - \tilde{\rho}_\mu(x) \,
[U^\dagger_\mu(x)]_{ba} \, \bar{\psi}_b(x+\hat{\mu}) \psi_a(x)$. Here,
$n_a(x)$, $p_{ab}(x,\mu)$, and $\tilde{p}_{ab}(x,\mu)$ are the dual
variables, which are either $0$ or $1$, and $a$ and $b$ are color
indices. The QCD partition function is then rewritten in terms of a
partition sum over the dual variables. Performing the Grassmann
integrals yields constraints of the form
\begin{equation}
\label{eq:constraints}
 n_a(x) + \sum_{\mu,b} \big[ \tilde{p}_{ba}(x-\hat{\mu},\mu) +
p_{ab}(x,\mu) \big] = n_a(x) + \sum_{\mu,b} \big[ p_{ba}(x-\hat{\mu},\mu) +
\tilde{p}_{ab}(x,\mu) \big] = 1 \,.
\end{equation}
In the final step one can perform the integrals over the gluonic
degrees of freedom, which is possible because they appear linearly in
the exponent.  Note that link variables also appear in front of the
exponential due to the terms in~\eqref{eq:ferm-act-exp}.  They can be
expressed as derivatives of the integral~\eqref{eq:I-int}. 

This procedure is not optimal in the sense that gauge invariance is
not manifest because the constraints \eqref{eq:constraints} are
incomplete.  Using only \eqref{eq:constraints} one would generate many
configurations whose contribution to gauge-invariant quantities is
zero after averaging over gauge fields.  We are currently deriving
additional constraints on the dual variables to ensure that we sample
only configurations whose contribution is nonzero after averaging.
This will lead to a partition function with similar constituents as
the partition function previously obtained in the infinite-coupling
limit, see, e.g., \cite{deForcrand:2014tha} and references therein,
except that we will have additional loops with contributions from the
auxiliary bosonic fields.  The resulting partition function will
probably have a sign problem and therefore needs further treatment to
make numerical simulations possible.

\section{Conclusions}

We have presented a study of the continuum limit of a theory that is
conjectured to provide an alternative discretization of Yang-Mills
theory. Numerical results show good agreement with simulations using
the usual Wilson action, for quantities at both zero and nonzero
temperature, already at finite lattice spacings in the approach to the
continuum limit. This provides evidence that the continuum limit
indeed corresponds to Yang-Mills theory. The new theory, when
formulated in terms of auxiliary bosonic fields, can be used to derive
dual presentations of QCD by methods previously only applicable in the
infinite-coupling limit.  Further details will be given in our future
publication~\cite{future-paper}.

\acknowledgments

We thank Robert Lohmayer for collaboration on the project and
Christoph Lehner for contributions at an early stage. We also
acknowledge very useful discussions with Philippe de Forcrand and
H\'elvio Vairinhos on their work~\cite{Vairinhos:2014uxa}.

\bibliographystyle{JBJHEP_mod}
\bibliography{lat14}

\end{document}